\RequirePackage{amsmath}
\documentclass{llncs}
\usepackage{times}         
\usepackage{type1cm}        
\usepackage{amssymb}
\usepackage{mathtools}
\usepackage{amsfonts}
\usepackage{graphicx}        
\usepackage{tikz}
\tikzstyle{densely dotted}= [dash pattern=on \pgflinewidth off 1pt]
\usepackage{multicol}        
\usepackage[bottom]{footmisc}
\usepackage{algorithm,algorithmic}
\usepackage{nth}
\usepackage[hidelinks]{hyperref}
\usepackage{booktabs}
\usepackage{helvet}  
\usepackage{courier}  
\usepackage{url}  
\usepackage{bm}
\input{xy}\xyoption{curve}

\usepackage[utf8]{inputenc}

\definecolor{red}{rgb}{1,0,0}
\definecolor{blue}{rgb}{0,0,1}

\begin{document}
\title{Dyslexia and Dysgraphia prediction:\\A new machine learning approach}
\author{Gilles Richard\inst{1} \and Mathieu Serrurier\inst{1} }
\institute{Dystech, Traralgon, Australia\\
\email{\{gillesr,mathieus\}@dystech.com.au}
}
\maketitle

\begin{abstract}
Learning disabilities like dysgraphia, dyslexia, dyspraxia, etc. interfere with academic achievements but have also long terms consequences beyond the academic time. It is widely admitted that between 5\% to 10\% of the world population is subject to this kind of disabilities. 
For assessing such disabilities in early childhood, children have to solve a battery of tests. Human experts score these tests, and decide whether the children require specific education strategy on the basis of their marks.
The assessment can be lengthy, costly and emotionally painful. 
In this paper, we investigate how Artificial Intelligence can help in automating this assessment. 
Gathering a dataset of handwritten text pictures and audio recordings, both from standard children and from dyslexic and/or dysgraphic children, we apply machine learning techniques for classification in order to analyze the differences between dyslexic/dysgraphic and standard readers/writers and to build a model. The model is trained on simple features obtained by analysing the pictures and the audio files.
Our preliminary implementation shows relatively high performances on the dataset we have used. 
This suggests the possibility to screen dyslexia and dysgraphia via non-invasive methods in an accurate way as soon as enough data are available.
\end{abstract}

\section{Introduction}
“Learning Disabilities” (or learning disorders) is an “umbrella” term describing a number of specific disorders, such as dyslexia, dyspraxia, dysgraphia, etc. A detailed view can be found in the book  {\it Diagnostic and Statistical Manual of Mental Disorders}  \cite{dsm5} (also known as DSM-5) from the American Psychiatric Association.
Diverse studies suggest that learning disabilities are characterized by subtle and spatially distributed variations in brain anatomy. As such, they should not be confused with learning problems which could be the result of visual, hearing, or motor handicaps, or even social issues.

Clarifying the neurological underpinning of a learning disability  has been a serious goal of research over the past twenty years.  Despite much progress has been made across diverse research fields, learning disabilities causes are still not well understood.
These learning disorders can be investigated from a lot of other viewpoints as well, such as:
\begin{itemize}
    \item Their prevalence depending of the gender \cite{Quinn2013}, 
    \item Their prevalence depending of family \cite{Snowling2016},
    \item The impact of alternative operational definitions of 'learning disabilities'  \cite{Fuchs2004,VanIJ1994,Waesche2011}
\end{itemize}
Nevertheless, we can agree with \cite{Tamboer2014} who thinks that: \emph{ it is possible to identify dyslexia with a high reliability, although the exact nature of dyslexia is still unknown}. We consider this is also valid with other learning disorders.
Despite a lack of understanding of the causes, the symptoms are generally clear and described in a comprehensive way in the DSM–5 under the terms of {\it Developmental Coordination Disorders}. To make it short:
\begin{itemize}
    \item Dyslexia is a learning disorder which impacts the individual ability to read.
    \item Motor dysgraphia is a learning disorder which impacts the individual ability to write.  
\end{itemize}
As suggested in \cite{lopez2018developmental}, motor dysgraphia 
might also be a marker of Developmental Coordination Disorder (DCD) such as dyspraxia.
These disorders remain over the age but can be mitigated with appropriate training sessions.
Obviously, it is not a matter of Yes or No and the symptoms range from mild to severe.
It is widely admitted that more or less 10\% of the world population has (to some extend) a learning disorder. One can refer for instance to \cite{DukeReport2016} from Duke University or again to DSM-5. 
It is well known that any combination of the disorders described in DSM-5 often leads to academic failure.
Nevertheless, provided with an appropriate education strategy, a child with learning difficulties will acquire the same skills as a standard child. 
Very often, these children can also get government supports in diverse ways (specific teaching lessons, extra tuition, extra time for exams, specific staff helping during the classroom,  etc.). To get such a support, the criteria is to provide a certificate coming from an accredited specialist, who is in charge of assessing the child.
The assessment may be lengthy, costly and emotionally painful.  Moreover
the limited number of accredited pathologists may make this process time consuming.
This situation obviously prevents a lot of people, often among the most targeted population, to carry on an assessment. This makes the search of fast, effective and widely available assessments (or pre-assessments) of primary interest. Our approach is typically a candidate solution to this issue. 

In fact, there has been an increasing interest to carry over the success stories of machine learning in diverse domains, from natural language understanding for chat bots to cancer detection. We feel it makes sense to investigate machine learning techniques to assess dyslexia and dysgraphia from :
\begin{itemize}
\item Writing: a picture of handwritten text  for motor dysgraphia 
\item Reading: a relatively small number of audio recordings for dyslexia
\end{itemize}
Our working hypothesis is that a properly trained machine learning algorithm might be able to distinguish between standard children and children with dyslexia and/or dysgraphia from a set of simple features. In the case of dyslexia we consider features extracted from 32 audio files of word reading (known words and non-sense words) per people. For dysgraphia, we use only features based on a picture of handwritten text (one per individual). 
We have experimented this approach on a restricted set of data coming from speech pathologists partners in Australia. Our  preliminary results are promising. 
This suggests that a screening could be implemented, providing a simple and accurate solution available for a large population at limited cost. 

This paper is organized as follows.
In Section \ref{summary}, we provide a very brief review of the machine learning process, explaining what we need in order to build a successful predictor. In Sections \ref{dyslexiaprinciple} and \ref{dysgraphiaprinciple}, we describe the main principles underlying our method for dyslexia and dysgraphia.
Section \ref{expe} is dedicated to our experiments: we describe the dataset, the protocol and the results we get. In Section \ref{related}, we review some related works, essentially for dyslexia as dysgraphia has not been widely investigated so far. We provide future works in our conclusion Section \ref{conclusion}.

\section{Machine learning: (very) brief summary}\label{summary}
Machine Learning (ML) is a sub-field of Artificial Intelligence which has been very successful over the past 10  years. The main idea is to start from a set (the bigger the better) of data and to try to automatically extract patterns or features which allow to provide a conclusion for unknown new data (see \cite{Nielsen2015} for a good introduction).
With the progress in the field of neural networks, it is today possible to build a predictor (for car or bicycle) which has an accuracy of 98\% i.e. its prediction is right in 98\% of the cases.\\
The 'car' example  seems basic but it is exactly the same as taking a picture of a small protuberance on your skin and asking 
'Is it a skin cancer ?' Today, we are able to build a predictor more accurate than any human expert in terms of skin cancer detection (see for instance \cite{Esteva2017} from Stanford University). Obviously, there is a huge mathematical background behind the scene, from statistics to complexity theory, through convex optimization. Building upon these theoretical achievements, powerful libraries have been developed allowing to abstract from the mathematical details, to design and implement clever algorithms leading to very accurate predictors. This is the core theoretical and technical framework underpinning our work in this paper.

\section{Dyslexia prediction principle}\label{dyslexiaprinciple}
One of the main symptoms of dyslexia is difficulty in reading. Our idea is then to gather reading audio recordings, from both dyslexic and non dyslexic readers, then to apply a machine learning algorithm. We expect the algorithm will {\it learn} the hidden characteristics allowing
to distinguish between dyslexic and non dyslexic people. Let us start with what a user is supposed to produce.
\subsection{Words selection and generation}
Our process is to have every child to read 32 words (no sentences, only words).
It is also well-known that dyslexic children struggle when it comes to reading words they have never seen or heard. 
They have also difficulties with some letters, or combination of letters ($p$ and $q$ for instance) and syllables.
Our initial corpus is coming from a set of 82 children’s books from the Gutenberg Project  \cite{gut}. We clean the texts and remove proper nouns. We obtain a list of around 100 000 words. Then we produce two lists : one with words from 4 to 6 letters,  one with words from 7 to 9 letters. In each list, we consider only words with high frequency of apparitions to guarantee  the words are known from children. After filtering, each of the two lists contains around 2000 words.

In a second step, we create two lists of pseudo-words which do not belong to the English dictionary but which {\it look like} English words. We also need the guarantee that the pseudo-word is pronounceable. In order to achieve that, we build a Long-Short Term Memory neural networks (LSTM) \cite{Hoch1997} that learn to build such pseudo-words. We are then able to generate an infinite list of pseudo-words. As for the real words, we build two lists of pseudo-words with different size and we keep only pseudo-words that fit with the following constraints :
\begin{itemize}
\item The word is not in the English dictionary
\item Every subset of 4 consecutive letters exists in an English word (to guarantee the word is pronounceable)
 \item It contains difficult letters or difficult combination of letters for dyslexic people.
\end{itemize}
The final list of 32 words to be read by a child is obtained by choosing 16 words in the list of real words and 16 words in the list of pseudo-words.
We change the length of the words with respect to the age of the user that performs the assessment :
\begin{itemize}
\item List 1: From 6 to 8 years old (included) the list is ordered this way:
\begin{itemize}
\item 2 easy real words
\item $\frac{2}{3}$ of words from  4 to 6 letters (50\% real, 50\% pseudo):
\item $\frac{1}{3}$ of words from  7 to 9 letters (50\% real, 50\% pseudo):
\end{itemize} 
\item List 2: From 9 to 13 years old (included) the list is ordered this way:
\begin{itemize}
\item 2 easy real words
\item $\frac{1}{3}$ of words from  4 to 6 letters (50\% real, 50\% pseudo):
\item $\frac{2}{3}$ of words from  7 to 9 letters (50\% real, 50\% pseudo):
\end{itemize} 
\item List 3: 14 years old and over the list is ordered this way:
\begin{itemize}
\item 2 easy real words
\item 100\% of words from  7 to 9 letters (50\% real, 50\% pseudo):
\end{itemize} 
\end{itemize} 
These constrained lists of words are randomly generated and are age-related: short words with simple syllables for children from 7 to 8, more difficult
for children from 9 to 13, then difficult for children over 14.
\subsection{Dyslexia input parameters}
For every word audio record, we consider 3 parameters:
\begin{itemize}
    \item The fact that the word (real or nonsense) has been properly pronounced according to the English pronunciation rules. This is a binary parameter (Yes or No).
    \item The fact that the word has been read one shot or the user, facing a difficulty, gets back to the beginning.
    This is still a binary parameter (Yes or No).
    \item The reading time measuring the interval between the display of the word and the reading start. The time unit is millisecond (ms). It is a real-valued parameter.
\end{itemize}
Each record is manually tagged (with a dedicated in-house software) except the reaction time which is evaluated by the computer.

\section{Dysgraphia prediction principle}\label{dysgraphiaprinciple}
As pointed out in our introduction, dysgraphia can be observed from a handwritten document and interferes at diverse levels of the handwriting process:
\begin{itemize}
    \item formation of the letters,
    \item letter size regularity,
    \item ability to follow a straight line, 
    \item gap between the average size of x-height letters and ascending/descending letters,
    \item etc.
\end{itemize}
\subsection{Dysgraphia input parameters}
\begin{figure}[!ht]
    \centering
    \includegraphics[width=1\linewidth]{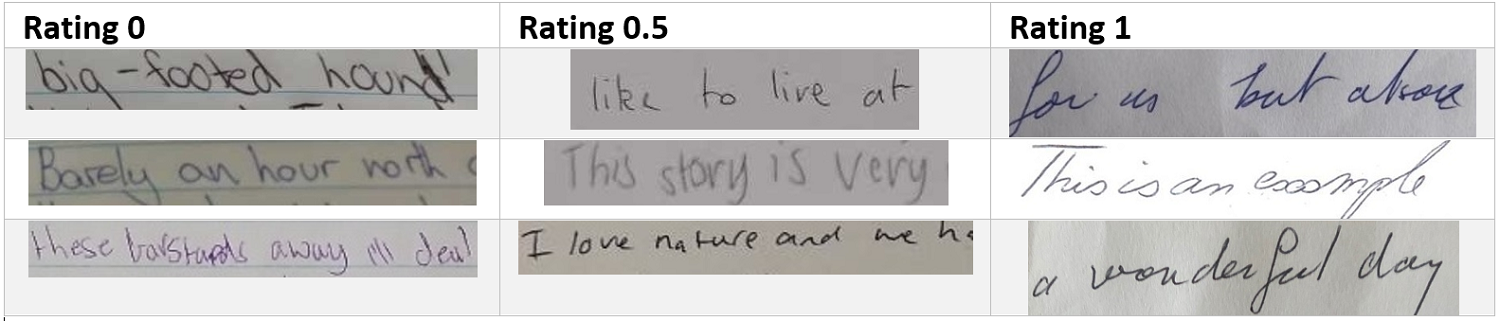}
    \caption{An example of ratings for slant}
    \label{slantrating}
\end{figure}

\begin{figure}[!ht]
    \centering
    \includegraphics[width=1\linewidth]{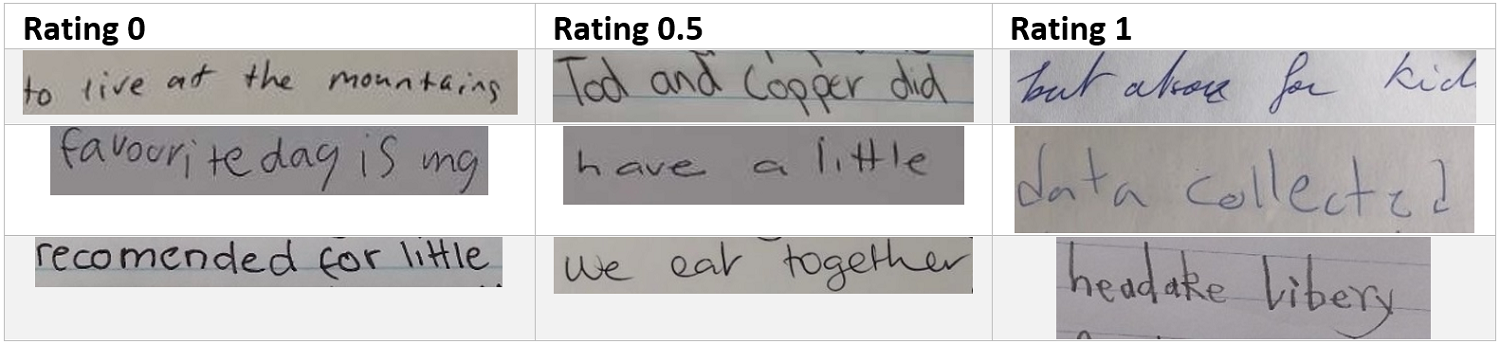}
    \caption{An example of ratings for amplitude}
    \label{amprating}
\end{figure}

\begin{figure}[!ht]
    \centering
    \includegraphics[width=1\linewidth]{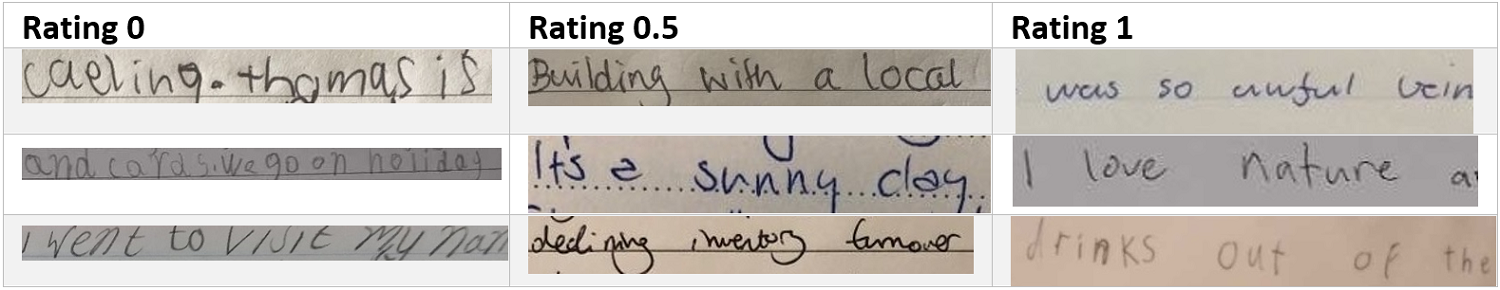}
    \caption{An example of ratings for word spacing}
    \label{spacingrating}
\end{figure}

In order to train our algorithm, we need a set of labelled data. We consider only pictures with at least 4 lines of handwritten text. It is also required to have only written text and no other signs on the picture. 
To ensure the network to be as robust as possible, the analysis has to work in various situations: the text is free, the paper can have lines or not, the handwritten text doesn't have to be centered, we can have scribbles, etc.. The text has to be written with Roman alphabet but apart from this limitation, the language is free. As usual in machine learning, the more diverse the data, the better the prediction.
The output of the algorithm is not only a likelihood of dysgraphia, but also an analysis of the handwriting features. These features have two interests at least. First, it will help to explain the prediction. The difficulty of providing explanations for a prediction is a major issue in machine learning. By providing these features, we can check if the features are realistic (most of them are easily understandable) and are consistent with the likelihood estimation. Second, as we will see in the next section, these features will also help for the training process. The features we consider are listed below:
\begin{enumerate}
    \item {\it Slant} : The slant feature corresponds to the direction of the handwriting. We normalize by converting the values from 0 (left slant) to 1 (right slant).
    \item {\it Pressure}: The estimated pressure of the handwriting from 0 (low)  to 1 (high).
    \item {\it Amplitude} : This is the average gap size between x-height and ascending/descending letters, from 0 (low) to 1 (high).
    \item {\it Letter Spacing}: This estimates the average spacing between letters in a word. Typically, a cursive writing style will lead to 0, from 0 (small spacing0 to 1 (large spacing).
    \item {\it Word spacing}: This estimates the average spacing between words in a sentence, from 0 (small spacing) to 1 (large spacing).
    \item {\it Slant Regularity}: from 0 (not regular) to 1 (highly regular). 
    \item {\it Size Regularity}: from 0 (not regular) to 1 (highly regular). Measure if the same letters vary.
    \item {\it Horizontal Regularity}: from 0 (the text doesn't follow an horizontal line) to 1 (the text follows an horizontal line). 
\end{enumerate}
Figures \ref{slantrating}, \ref{amprating} and \ref{spacingrating} represents example of respectively slant, amplitude and word spacing rating: rating 0 on the left side, 0.5 on the middle and 1 on the right side.

Currently, all features are manually labelled with a dedicated in-house software. In the medium terms, this  process will be done automatically via machine learning again. 

\section{Experiments}\label{expe}
\subsection{Context and metrics}
Machine learning is a data-driven technology. Apart from designing an algorithm, we have to gather proper data, with their own labels. Not only the quality, but also the quantity is important as training neural networks requires a lot of data in order to be accurate.
In \cite{wagner2018}, the author points out that the difficulty to diagnose dyslexia is mainly coming from the unbalanced population. Only a relatively low percentage of the population has dyslexia or dysgraphia, and so to train a mathematical model with a so unbalanced population is always a challenge. When it comes to measure the performance of the algorithm, standard accuracy can then be a misleading metric. Assuming we have 10\% of the population having dyslexia or dysgraphia, a baseline algorithm declaring everybody as non dyslexic/dysgraphic will ensures a 90\% accuracy. 
As a consequence, other metrics are needed like precision, recall and F1-score. Let us recall below their classical definitions. For a binary classifier, we have at our disposal a set of positive examples (dyslexic children) and a set of negative examples (non dyslexic children).
We note $tp$ the number of positive examples predicted as positive, $tn$ the number of negative examples predicted as negative, $fp$ the number of negative examples predicted as positive (false positive), and $fn$ the number of positive examples predicted as negative (false negative). The metrics are defined as follows:

$$
accuracy=\frac{tp+tn}{tp+tn+fp+fn}
$$
The accuracy measures the probability that the class predicted by the model is the right one. In the latter we will use percentage for this accuracy.
$$
precision=\frac{tp}{tp+fp}
$$
The precision is the probability of being positive if the example is predicted as positive. In some sense, this measures the correctness of the predictor when it predicts an example as positive. The bigger this number, the better the predictor is.
$$
recall=\frac{tp}{tp+fn}
$$
The recall is the probability of a positive example to be predicted as positive. In some sense, this measures the ability of the predictor to predict all positive example as positive. Still, the bigger this number, the better the predictor is.
$$
f1\text{-}score=2*\frac{precision*recall}{precision+recall}
$$
The f1-score is a balance between precision and recall. Thus, accuracy focus on the performances of the model in general when precision, recall and f1-score focus on performances of the model on positive examples only.\\

We compare the performances of four state of the art classifiers: Dummy classifier that chooses classes randomly with the a priori probability of classes computed on the training set, Naive Bayes classifier, logistic classifier and random forest. In order to have average estimation of the metrics previously described, we use a $k$-fold cross-validation scheme for each experiment (5-folds for dyslexia and 10 folds for dysgraphia).

\subsection{Dyslexia results}
\begin{table}[ht!]
\centering
\label{dyslex-table}
\begin{tabular}{|l|c|c|c|c|}
\hline
\textbf{Alg.} & Accuracy & Precision & Recall & f1  \\ \hline 
Majority class &47.9 [11.0] & 0.58 [0.12] & 0.48 [0.11] & 0.52 [0.11]\\ \hline \hline 
Naïve Bayes & 89.6 [7.3] & 0.94 [0.06] & 0.88 [0.07] & 0.91 [0.06]\\ \hline
Logistic reg. & 87.1 [9.5] & 0.92 [0.12] & 0.88 [0.11] & 0.89 [0.08]  \\ \hline
Random Forest & 90.0 [5.7] & 0.95 [0.06] & 0.88 [0.07] & 0.91 [0.05]\\ \hline
\end{tabular}
\vspace{0.2cm}
\caption{Performance of machine learning algorithm for detecting dyslexia with 5-fold cross-validation}
\end{table}
As pointed out in Section 3, we have developed a mobile app (Android and IOS) enabling professionals, specialist clinics and schools to participate in our research by helping us collecting data. The experiment is done on $69$ people, including $41$ peoples with official dyslexia diagnostic. For each individual, we get $32$ audio files that are manually tagged. Thus, for each session we obtain the following 10 features:
\begin{itemize}
    \item age in year
    \item average error
    \item average backtracking
    \item average reading time
    \item average error for English words
    \item average backtracking for English words
    \item average reading time for English words
    \item average error for generated words
    \item average backtracking for generated words
    \item average reading time for generated words
\end{itemize}
The average values of metrics (with standard deviation in brackets) are described in Table \ref{dyslex-table}. We can observe that machine learning algorithm clearly outperforms random guess. The best results are obtained with Random Forest which achieve $90\%$ of accuracy. The precision of $0.95$ indicates that when the models predict the class dyslexic, it is right $95\%$ of the time. The recall indicates that it detect $88\%$ of the dyslexic people in average. These results could be close to optimal for two reasons : i) most of non-dyslexic individuals have never passed a dyslexia test with a clinician, ii) some of dyslexic individuals in the dataset attend training sessions to lower the symptoms of dyslexia. Thus, there is some noise in the dataset which could be partially overcome by considering more data. In this case, we can reasonably expect an higher accuracy.
\begin{figure}[!htb]
\centering
\begin{minipage}{.5\textwidth}
  \centering
  \includegraphics[width=1\linewidth]{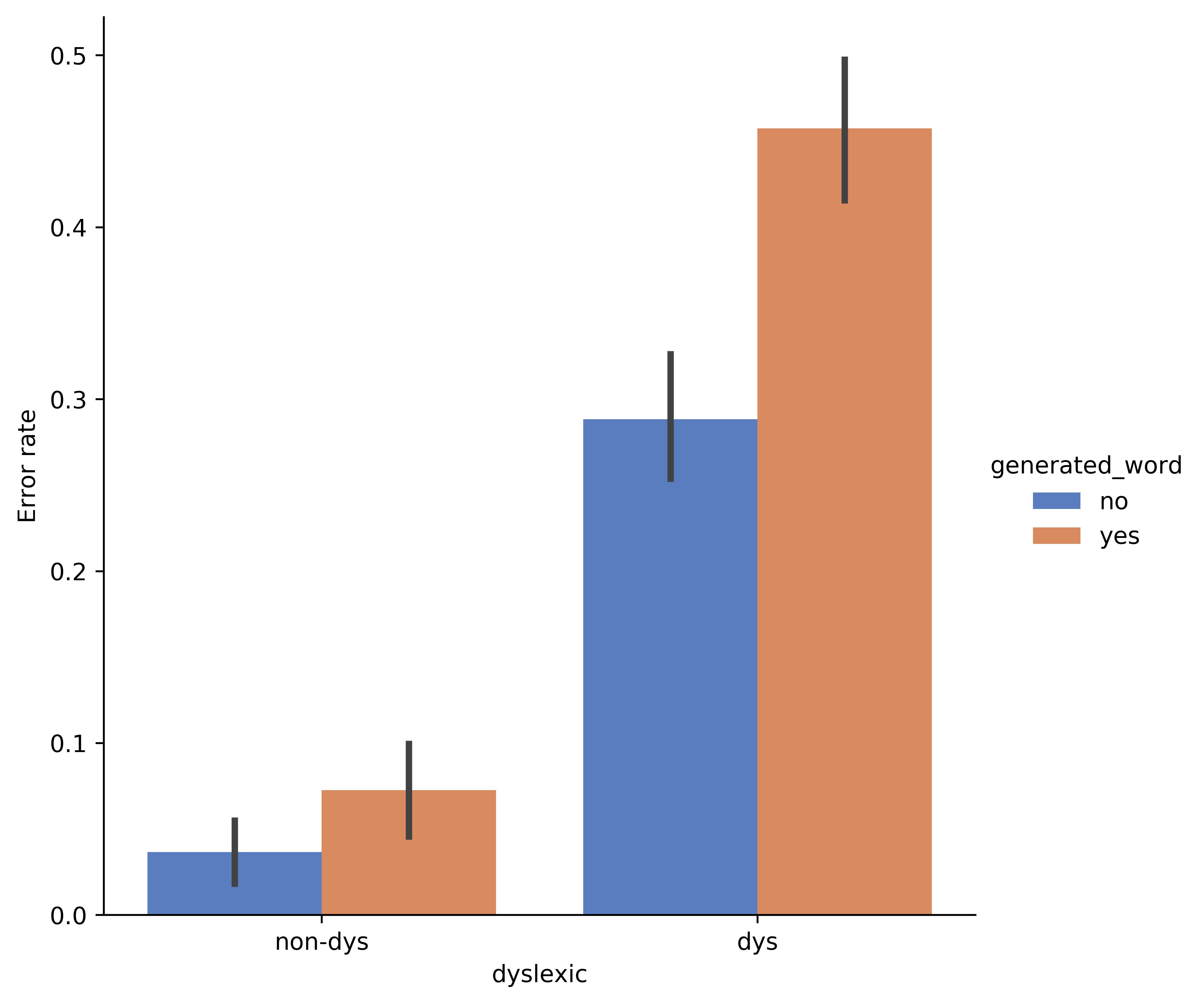}
  \caption{Error rate for non-dyslexic/dyslexic person with respect to type of word}
  \label{fig:test1}
 \end{minipage}\hfill
\begin{minipage}{.5\textwidth}
  \centering
  \includegraphics[width=1\linewidth]{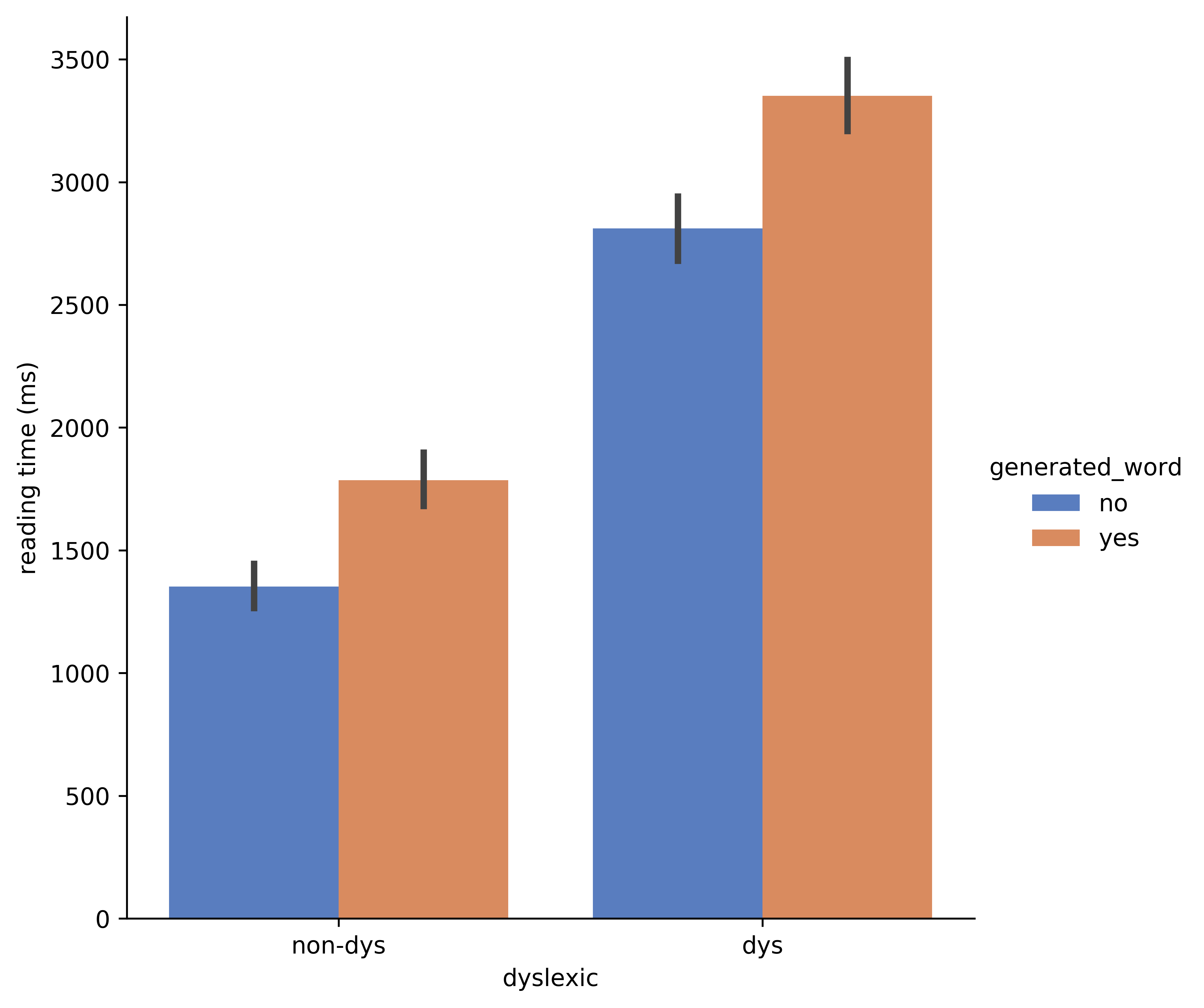}
  \caption{Reading time for non-dyslexic/dyslexic person with respect to type of word}
  \label{fig:test2}
\end{minipage}
\end{figure}
The results suggest that the simple test we propose is effective for detecting dyslexia. Especially Figure \ref{fig:test1} shows that, as expected, the error rate is significantly higher for dyslexic people. It also demonstrate the generated words are far more difficult to read for dyslexic people than for non-dyslexic one. This difference also appears significantly when considering the reading time (Figure \ref{fig:test2}). 

\subsection{Dysgraphia results}
Our dataset is composed of 1481 pictures of free handwritten text. Each text has at least five lines of text. The dataset contains 198 pictures of peoples diagnosed with dysgraphia, about 13\% of the dataset.
For each picture of the dataset, the 9 features has been reported manually by the same person. In order to avoid biases, this person was not aware of the diagnostic associated to the picture. 
In that case, we are also in a position to predict the features values.

\begin{figure}[!ht]
    \centering
    \includegraphics[scale=0.7]{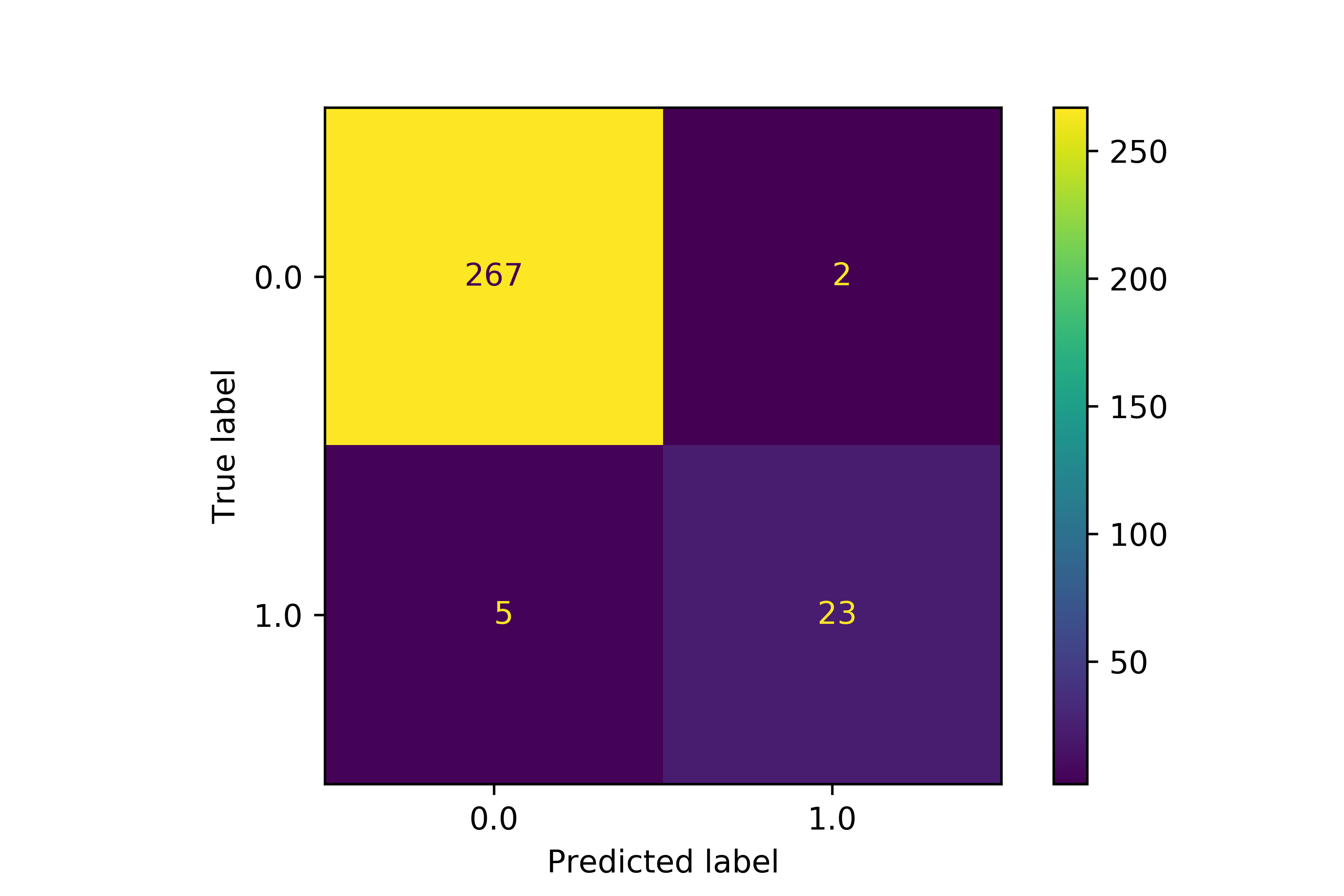}
    \caption{Confusion matrix for dysgraphia prediction on a test set of 297 examples}
    \label{confmat}
\end{figure}

\begin{table}[ht!]
\centering
\label{dysgraph-table}
\begin{tabular}{|l|c|c|c|c|}
\hline
\textbf{Alg.} & Accuracy & Precision & Recall & f1  \\ \hline 
Majority class &74.7 [1.2] & 0.10 [0.04] & 0.11 [0.06] & 0.10 [0.05] \\ \hline \hline 
Naïve Bayes & 90.8 [2.9] & 0.62 [0.08] & 0.93 [0.11] & 0.73 [0.06]\\ \hline
Logistic reg. & 95.6 [2.9] & 0.87 [0.12] & 0.82 [0.19] & 0.82 [0.14] \\ \hline
Random Forest & 96.2 [2.7] & 0.92 [0.08] & 0.78 [0.19] & 0.83 [0.14]  \\ \hline
\end{tabular}
\vspace{0.2cm}
\caption{Performance of machine learning algorithm for detecting dysgraphia with 10-fold cross-validation}
\end{table}
The metrics for the dysgraphia prediction for the four algorithms are presented in Table \ref{dysgraph-table}. As for dyslexia, the best model is Random forests with an accuracy of $96.2\%$ (random guess achieves $74.7\%$). The precision and recall are quite high (resp $0.92$ and $0.78$ against $0.10$ and $0.11$ for random guess). Since the classes are unbalanced, these last scores are more significant than accuracy. The confusion matrix for random forests on a test set is presented in Figure \ref{confmat}. This shows dysgraphia can be predicted with high accuracy from a simple analysis of an handwritten text. Once again, we can assume than the classes are noisy and the accuracy could be improved by collecting more data with official diagnostics. 

\section{Related works}\label{related}
So far, much more efforts have been deployed in the field of dyslexia detection than for dysgraphia.
In \cite{Frid2018}, the authors start from the hypothesis (Asynchrony Theory) that dyslexia could come from a gap in the speed of processing between the different brain entities activated in the word decoding process. This gap may prevent the synchronization of information necessary for an accurate reading process. Starting from this, they monitor a population of 32 children, with a more or less 50/50 percentage of dyslexic/standard readers. At this stage, we do not know how the authors balance the fact that their dataset does not match the 90/10 usual split of population between dyslexic and standard children.

Making the children to read 96 real words and 96 non-sense words, they record brain activity via electroencephalogram (EGG) and implement a binary classification algorithm (namely Support Vector Machines SVM) to distinguish between dyslexic and standard readers. They obtain an accuracy of $78.5\%$, with a precision equal to $0.78$, a recall equals to $0.78$ and f1-score equals to $0.78$.
Apart from the fact that our population is lesser than in \cite{Frid2018}, it is clear that our technology is far less invasive than using EGG and obtain better results. 
As such, our approach seems a more realistic approach when it comes to implement real applications.

Following the same idea of investigating brain activity, but in that case using neuro-imaging scans of the brain, \cite{Tamboer2016} start from a population of 49 students in a 50/50 split dyslexic/standard. The authors still implement a SVM classifier to get the following result : accuracy of $79.5\%$, with a precision equal to $0.75$, a recall equals to $0.81$ and f1-score equals to $0.78$.

After training the SVM on 49 data, they tested on a another dataset made of 876 subjects whose only 60 (7\%) where dyslexic. Then the accuracy goes down to $59.1\%$ with a with a precision equal to $0.10$, a recall equals to $0.66$ and f1-score equals to $0.18$. 


On view of the numerical results, it seems that their studies provide more support for the use of machine learning in anatomical brain imaging than for a realistic non invasive dyslexia screening application. We can also cite the works of \cite{Rello2015} where for the first time, an eye tracking technology associated to an SVM classifier was used to predict dyslexia starting from a dataset of 97 subjects, 48 of then with diagnosed dyslexia. The eye tracking technology allows to extract information such as {\it Number of visits} (Total number of visits to the area of interest in the text), {\it Mean of visit} (Duration of each individual visit within the area of interest in the text), etc. The resulting accuracy is in the range of 80\% which is quite good with regard to the 50/50 split of the dataset. Nevertheless, this could be far from the ground truth knowing that we have far less than 50\% of dyslexic in the real world. More recently, we can  cite the works of \cite{Asvesto2019}, still using eye tracking associated to an SVM algorithm and getting a very good 97\% accuracy rate over a set of 69 children, 32 being dyslexic. Their system {\it DysLexML} could ultimately be the basis of a screening tool as soon as the cost of eye-trackers allows to reach  a larger  population. 

\cite{Couso2010} start from a different hypothesis. Considering that the available datasets coming from speech pathologists contain uncertain information, they rely on a mix of fuzzy set theory and genetic algorithms to rigorously aggregate the data and deal with uncertainty. Their dataset (available on KEEL project web site http://www.keel.es) is coming from 65 children. It is not easy to compare with the other approaches as they use more than 2 classes in their result and the metrics used in fuzzy logic are not the standard ones. Nevertheless, their algorithm is part of a web-based, automated pre-screening application that can be used by parents.

\section{Future works and conclusion}\label{conclusion}
In this paper, we show that it is possible to predict dyslexia and dyspraxia based on very simple tasks by using Machine Learning technologies. It has been recently proved that properly trained ML-based predictors can be more accurate than human experts on specific task. Based on these facts and our encouraging results, we think there is a huge potential using machine learning to help people with learning disorders. As usual with machine learning, accuracy can still be improved by gathering more data. In the same way, machine learning algorithms, and particularly deep learning, could be used for processing pictures and audio data with minimal pre-processing. This would avoid the need of manual analysis and the global performances may also be improved. From another viewpoint and regarding dysgraphia, it is likely that it can be detected on a drawing only. If we are able to do that, this will make a screening for dysgraphia available for children under the age of writing. In the future, a better understanding of the correlation between the different disorders could also help in providing accurate predictions. This knowledge could come from the cognitive science community.
\bibliographystyle{plain}
\bibliography{biblio}
\end{document}